\documentclass[10pt,conference]{IEEEtran}
\IEEEoverridecommandlockouts
\usepackage{cite}
\usepackage{tcolorbox}
\usepackage{amsmath,amssymb,amsfonts}
\usepackage{algorithmic}
\usepackage{graphicx}
\usepackage{textcomp}
\usepackage{AMMALanguages}
\usepackage{xcolor}
\usepackage[utf8]{inputenc}
\usepackage{float}
\usepackage{comment}
\def\BibTeX{{\rm B\kern-.05em{\sc i\kern-.025em b}\kern-.08em
    T\kern-.1667em\lower.7ex\hbox{E}\kern-.125emX}}

\usepackage{ifthen}
\usepackage{amssymb}
\newboolean{showcomments}
\setboolean{showcomments}{true} % toggle to show or hide comments
\ifthenelse{\boolean{showcomments}}
  {\newcommand{\nb}[2]{
    \fcolorbox{gray}{yellow}{\bfseries\sffamily\scriptsize#1}
    {\sf\small$\blacktriangleright$\textit{#2}$\blacktriangleleft$}
   }
   
  }
  {\newcommand{\nb}[2]{}
   
  }

\begin{document}

%\title{Towards the usage of prompt learning methods for automated model completion \\
\title{Towards using Few-Shot Prompt Learning for Automating Model Completion %\lola{how about we remove ``prompt''? I have googled "few-shot prompt learning" and there are only 400 hits. I think it is more common to talk only about few-shot learning} \\ \houari{I think few shot prompt learning makes sense. Few shot learning is broader and does not necessary involve prompts.} \\

}

\author{
\IEEEauthorblockN{Meriem Ben Chaaben}
\IEEEauthorblockA{\textit{DIRO, Université de Montréal} \\ Montréal, Canada \\
{meriem.ben.chaaben@umontreal.ca}}
\and
\IEEEauthorblockN{Lola Burgue\~{n}o}
\IEEEauthorblockA{\textit{University of Malaga} \\
Malaga, Spain \\
lolaburgueno@uma.es}
\and
\IEEEauthorblockN{Houari Sahraoui}
\IEEEauthorblockA{\textit{DIRO, Université de Montréal} \\ Montréal, Canada \\
sahraouh@iro.umontreal.ca}
}

\maketitle

\begin{abstract}
We propose a simple yet a novel approach to improve completion in domain modeling activities. Our approach exploits the power of large language models by using few-shot prompt learning without the need to train or fine-tune those models with large datasets that are scarce in this field. We implemented our approach and tested it on the completion of static and dynamic domain diagrams. Our initial evaluation shows that such an approach is effective and can be integrated in different ways during the modeling activities.
\end{abstract}


\section{Introduction and Motivation}
%Focus on Prompt engineering / learning: how to move from NL to specific formalism. \\
%Intro and related work... \cite{RoccoSRN21}

%Prompt learning to move from the semantics of Natural Language (NL) to the semantics of a formalism. In our case, we illustrate it with class diagrams and activity diagrams...

Recent developments in deep learning-based language models (LMs) open a world of possibilities to automate and assist software specialists in software development and maintenance tasks.
At the implementation level, large code bases allow us to leverage these language models by pre-training them to have good code representations and by fine-tuning them on software engineering specific tasks.

These opportunities are, however, limited when dealing with early software development phases such as analysis and design. Datasets are scarce, and when available, they are not large enough to pre-train or fine-tune deep-learning models.
For software modeling activities, several contributions were proposed to circumvent the lack of large datasets. The goal of these research contributions is to recommend domain concepts, their features, and relationships during modeling activities\footnote{Note that ``deep-learning/language model'' and ``software model'' do not refer to the same type of model. To avoid confusion, in this work, each time we refer to a deep-learning or language model (LM), we always refer to it as such, and never as ``model'' alone.}.

Di Rocco et al.~\cite{RoccoSRN21} proposed an approach based on graph kernels that only need a small-size dataset for training. Although the results were promising, the quality of the recommended elements remains too low to be used in real settings.
Similarly, Weyssow et al.~\cite{WeyssowSS22}, used such model/metamodel datasets to train an LSTM neural network. Here again, the authors obtained limited results, especially when applied to the iterative construction of a metamodel.
From another perspective, Capuano et al.~\cite{Capuano22}, exploited the available large code bases to reverse-engineering a set of models to train a RoBERTa language model. The results are acceptable, but the  reverse-engineered models on which the authors had to rely for training led to suggestions that reflect implementation aspects rather than the modeled domains.
%Always with the perspective of exploiting other source of data,
With the goal of exploiting knowledge captured in general and specific natural-language documents, Burgueño et al.~\cite{Burgueno21} used these documents to train language models to suggest model completions. We believe that exploiting natural-language sources is a good idea to overcome the scarcity of the data to exploit deep-learning to assist in modeling activities. However, this work requires to train a language model from scratch for each specific domain, which remains a challenging problem in many scenarios.

\begin{figure}[h]
\begin{center} 
\includegraphics[width=0.45\textwidth]{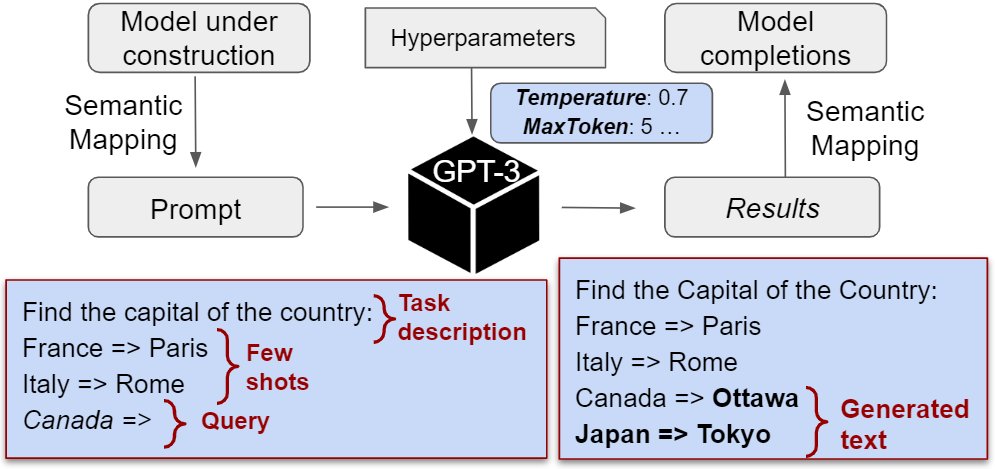}
\caption{Explanatory example}
\label{fig:conceptsQuery}
\vspace{-4mm}
\end{center}
\end{figure}

In this paper, we propose the novel idea of exploiting powerful left-to-right large language models (LLMs). %(Brown et al. NeurIPS 2020).
To this end, we use few-shot prompt learning, which allows us to exploit these LLMs without having to train or fine-tune them on a specific domain or task, and which has proven its efficiency to various natural language processing tasks~\cite{Liu22}. %(Liu et al. ACM Comp. Survey 2022).  This technique consists in giving very few examples (few shots) on the kind of information we are looking for and then give a specific text to complete (which we will call \emph{query}). 
For example, %In the following, we present a simple example where we explain these concepts. 
if the goal is to use a LLM to generate the name of the capital of a given country, we need to prepare a prompt where we first provide the LM a description on how to reply to our queries, then we add relevant labeled samples such as two countries and their corresponding capitals (two shots). Finally, we give the country for which we want the LM to provide its capital, i.e., autocomplete. Figure \ref{fig:conceptsQuery} gives an overview on how to use LLM with prompt learning, with the example of countries and capitals. Note that apart from being an autocomplete engine, these language models are also pattern matching and pattern generation engines. This is why we need to provide not only information about the task to perform but also the pattern that we want them to replicate.
%In our example, each prediction unit is a token or span and not an entire text, as a result it is recommended to put the max tokens parameters to a low value.
%The generated text will try to fit the pattern in accordance with how the prompt  is written.
As our example shows, the generated text could include further information. For instance, \texttt{Japan => Tokyo} has also been generated, which is not of our interest. To deal with this behaviour, properly transforming queries into prompts and results into modeling elements is an essential part of the approach.

We specifically take advantage of GPT-3, which is one of the most powerful LM (which contains 175 billion parameters~\cite{BrownMRSKDNSSAA20}), to represent most of the existing general concepts to support software specialists when modeling. To adapt the semantics of the general concepts represented in GPT-3 to the semantics of modeling formalisms, we use two semantic mappings: one that takes the model under construction (i.e., the input to our system) and builds the prompt, and another one that obtains model auto-completion suggestions from the text produced by the language model. These semantic mappings rely heavily on the targeted modeling formalisms.

In this new-ideas paper, we illustrate our approach with two examples coming from two categories of modeling languages: static models, i.e., UML class diagrams; and dynamic models, i.e., UML activity diagrams. We propose an initial implementation of this idea and a preliminary evaluation.

%Related work goes here, too.

\section{Prompt-Learning for Model Completion}

\subsection{Approach Overview}

%Our key focus is to leverage the powerful language model GPT-3   to assist in software engineering activities. \\

The main goal of our approach is to complete a model under-construction (a.k.a. partial model) by suggesting related elements.
Given a partial model,
%We first consider a query to our engine as a partial model. This is the starting point of our process.
we apply a semantic mapping, i.e, we construct a text representation that will serve as input to GPT-3. We query GPT-3 and it returns a textual output that follows a certain pattern. Then, we finish by applying another semantic mapping---in particular, a parsing---to the obtained text and extracting relevant model elements by applying suitable text transformations.

%With the above concepts in mind, we expand in our study, this underlying paradigm to do what is called Semantic Parsing \cite{Liu22}. It is an information extraction task where the goal is to produce a structured meaning representation from a natural language input. We study then in this work this proposed strategy and how far we are able to apply it  on model completion tasks.

\subsection{Static diagram completion}
\label{sec:staticDiagCompletion}

\begin{figure}[h]
\centering
\includegraphics[width=0.75\columnwidth]{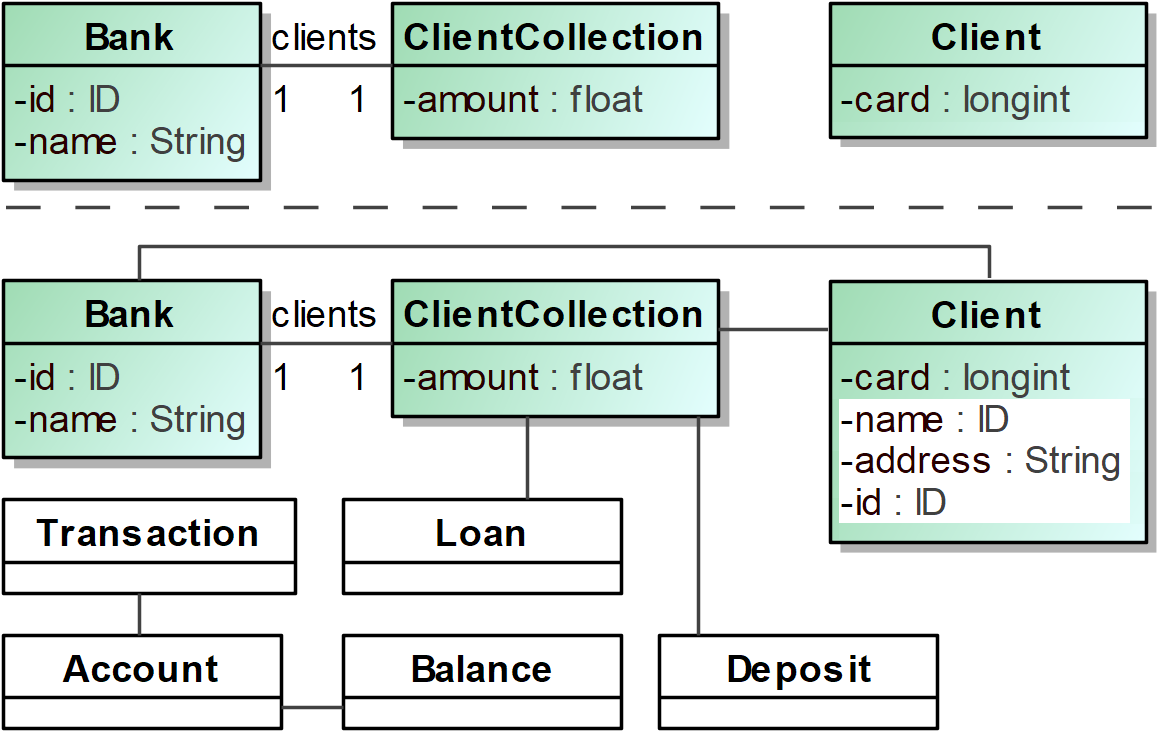}
\caption{Example of domain model: Bank class diagram}
%\lola{I am having second thoughts about presenting here the whole diagram (i..e, the ground truth) as it can be confusing. The comparison of our suggestions with the ground truth is the technique that we have used for our evaluation, but probably, in this section where we are presenting our approach, we shouldn't present the autocompletion of a model for which we already have the final model. This is, maybe we should only present the partial model inside the dotted box and say how we can complete it. What do you think? If you agree, I could do the corresponding changes in the figures and text.} \Meriem{I AGREE !}
\vspace{-4mm}
\label{fig:classDiagram}
\end{figure}

Using static diagrams for domain modeling usually consists of representing the domain entities, their properties or features and their relationships. For example, the UML class diagrams shown in Fig.~\ref{fig:classDiagram} is a partial description of a banking system. The upper part represents the partial domain model that is already defined by the user.

\begin{figure}[H]
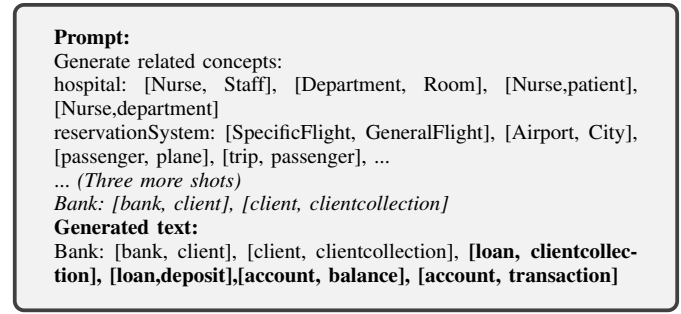

\begin{tcolorbox}
\footnotesize
\textbf{Prompt: }\\
Generate related concepts: \\
hospital: [Nurse, Staff], [Department, Room], [Nurse,patient], [Nurse,department] \\
reservationSystem: [SpecificFlight, GeneralFlight], [Airport, City], [passenger, plane], [trip, passenger], ...\\
    ... \emph{(Three more shots)}
%\lola{are we ommiting a lot here? is it feasible to add the complete shots? It would be a closer idea about the fact that not a lot of text is required for GPT-3 to understand what it is requested to do} \\
\\
\textit{Bank: [bank, client], [client, clientcollection]} \\
\textbf{Generated text: }\\
Bank: [bank, client], [client, clientcollection], \textbf{[loan, clientcollection], [loan,deposit],[account, balance], [account, transaction]}

\end{tcolorbox}
\caption{Prompt and generated text for class names and association prediction to complete the Class Diagram of Fig.~\ref{fig:classDiagram}}
\label{class-prompt}
\vspace{-4.5mm}
\end{figure}

We focus first on how we design our completion system to suggest entities, i.e., new classes.
We create the prompt using some existing diagrams of unrelated domains. From these diagrams, we extract pairs of related classes and, to follow a certain pattern where we introduce the relationship between two related elements, we represent them between brackets as shown in Fig.~\ref{class-prompt}. These are the few-shots that we provide. Then, we build a query from the partial domain model. To do this, we select between 2 and 4 pairs of related classes, put the classes names in brackets and add them to the prompt. In Fig.~\ref{class-prompt}, under ``Generated text'' we can observe the text that has been generated for our example when the model has been hyper-parameterized with a temperature value of 0.7 which is a setting  to control the randomness and creativity of the model’s predictions, and a maximum number of tokens of 20.
Furthermore, our engine follows a ranking strategy to suggest new elements. We query GPT-3 several times with different prompts, where all the prompt have the same shots but different queries, each query containing a different subset of model elements from the partial model. As a result, we obtain for each prompt a set of suggested concepts. Then, all the obtained concepts from the different prompts are ranked by its frequency from higher to lower. Only those concepts with higher frequency are considered.

We apply a string-searching algorithm on the generated text to extract relevant class names and the association that  exist between them; we also remove spelling errors and noisy data such as digits, which are usually not part of domain models. In our example, after this step, we obtain that potential missing classes (and associations between them) are \texttt{Transaction, Balance,Deposit, Account} and \texttt{Loan}. These classes are suggested to the user as shown in the bottom part of Fig.~\ref{fig:classDiagram} with a white color.
One can notice that for this example all the suggestions are closely related to the banking domain that is being modeled.

%\begin{figure}[h]
%\begin{center}
%\includegraphics[width=8cm]{figures/result.png}
%\caption{Genertaed model }
%\label{fig:resultsClassDiagram}
%\end{center}
%\end{figure}

%\Meriem{updated this in approach ..}
%For the remaining features (i.e., attribute and association suggestions), we follow the same steps as for new classes suggestions, but the semantic mappings that we use to generate the prompt and obtain the completions are different as we explain below.

Given a partial model, to generate prompts for attribute completion, we concatenate the package name and existing class names with their attributes in brackets, ending with the class for which we are finding potential attributes.
Like we do for classes, we use a frequency based ranking function that takes as input all the text generated by GPT-3 for different prompts. Then, we generate attribute suggestions using those concepts which are higher in the ranking.

Figure~\ref{fig:examplePromptClassDiagramAttributes} illustrates the prompt and resulting text for the class \texttt{Client} of Fig.~\ref{fig:classDiagram}. We have obtained \texttt{name, address} and \texttt{id} as potential relevant attributes.
\begin{figure}[h]
\begin{tcolorbox}
\footnotesize
\textbf{Prompt: }\\
      Generate missing attributes for each class in this class diagram:  \\
      package company: employee: [id, name, lastName, occupation]; manager: [id, name, department]; company: [name, holding] => employee: [id, name, lastName, occupation, department, experience, revenue]; manager: [id, name, department, team, revenue]; company: [name, holding, address, website] \\
      \textit{package bank: bank: [id,name]; clientCollection: [amount]; client: [card]} \\
      \textbf{Generated Text:}\\
      package bank: bank: [id, name]; clientCollection: [amount]; client: [card, \textbf{name, address, id]}

\end{tcolorbox}
\caption{Prompt and generated text for attribute completion.}
\label{fig:examplePromptClassDiagramAttributes}
\vspace{-2mm}
\end{figure}

Regarding the suggestion of association names, we design our prompt as follows; we introduce a few shots corresponding to pairs of classes and their association names that we select from unrelated diagrams. Then, we add both class names whose association name we are aiming suggest.
%\fbox{\parbox{\dimexpr\linewidth-2\fboxsep-2\fboxrule\relax}{\centering ['Bank', 'ClientCollection'] \\ [['ClientCollection','Client']] }}

%\lola{aren't we talking about associations in the end?}
%As we will show in the preliminary evaluation,
There are different ways of using the prompts for static diagram completion; either at each iteration of the modeling activity, one prompt at a time, or after the designer has completed his diagram to suggest potentially missing elements, with the combination of the results of many prompts.
\vspace{-1mm}
\subsection{Dynamic diagram completion}

\begin{figure}[h]
\begin{center}
\includegraphics[width=0.5\textwidth]{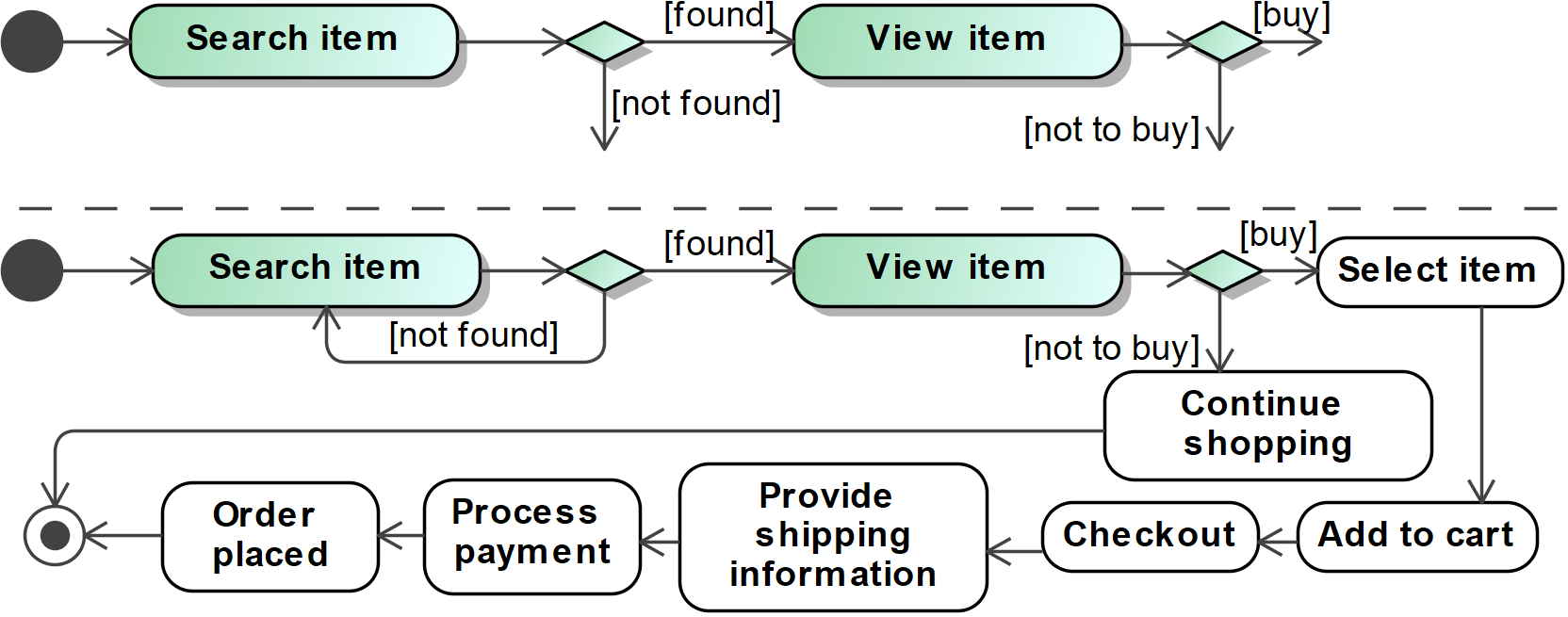}
\caption{Example of activity diagram: Online Shopping}
\label{fig:AD}
\vspace{-5mm}

\end{center}
\end{figure}

Since structural diagrams do not define sequences, any fragment can be used to generate the prompts to complete a diagram. In the case of dynamic/behavioral diagrams, such as activity diagrams~\cite{Rumbaugh2004}, there are strong precedence/sequence constraints to consider (e.g., to represent time), as it can be seen in Fig~\ref{fig:AD}. Hence, to apply prompt learning, we need to define shots and prompts in a way that preserves those constraints.

\begin{figure}[h]
\begin{center}
\includegraphics[width=0.5\textwidth]{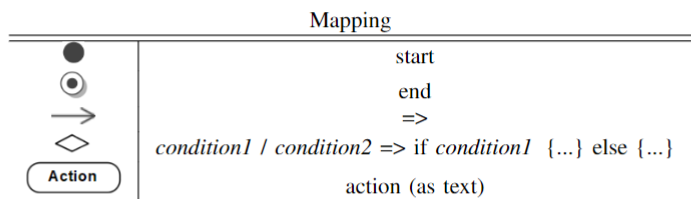}
\caption{Semantic Mapping for Activity Diagrams}
\label{fig:semanticMappingAD}
\end{center}
\end{figure}

Once again, we need to map the semantics of activity diagrams to a pattern that a LLM such as GPT-3 is able to understand and for which it provides meaningful results. To deal with precedence constraints, we designed our prompts and parameterised them to predict the next actions in a partial sequence. This implies to set the \texttt{max token} hyperparameter to a relatively low number---in our experiments it has been set to 50.
%When building the prompt, we have applied the  MDE principles and exploited the activity diagram metamodel by Rumbaugh et al.~\cite{Rumbaugh2004}.
To build the prompts, we defined simple transformation rules to match the elements of the activity diagram to the appropriate keywords that conform the prompt that we send to GPT-3 as Fig.~\ref{fig:semanticMappingAD} shows. Note that so far, we have only focused on a subset of the activity diagram language.

To illustrate our idea, we introduce the example of an online shopping workflow. Fig \ref{fig:AD}  shows, in the upper part, the partial activity diagram  that is already defined by the user.  \\
To create our prompt, we design 3 shots using real activity diagrams extracted from a public repository~\cite{OnlineAD}, which have been mapped using the rules described above. Fig~\ref{fig:examplePromptActivityDiagram} represents the prompt for this example and the GPT-3 generated text.
\begin{figure}[h]
\begin{tcolorbox}
\footnotesize
\textbf{Prompt:}\\
Complete the  workflow:

    Ticket Vending Machine: \textbf{start}  => request trip info => provide trip info => process trip info =>  payment request => provide payment Info => process payment => pay with card / pay with cash => \textbf{if} with card $\{$authorize card payment$\}$ => \textbf{end}.
    \\
    ... \emph{(two more shots)} \\
    \textit{Online Shopping: \textbf{start}  => search item => found / not found => \textbf{if} found $\{$view item => buy  /  not to buy$\}$\}=>}
    \\
\textbf{Generated Text:}

\textbf{if} buy $\{$select item => add to cart => checkout=> provide shipping information => process payment => order placed => \textbf{end}$\}$ \textbf{else} $\{$continue shopping => \textbf{end}$\}$\} \textbf{else} $\{$search$\}$.
\end{tcolorbox}
\caption{Prompt and generated text for the Activity Diagram of Fig.~\ref{fig:AD}}
\label{fig:examplePromptActivityDiagram}
\vspace{-3mm}
\end{figure}

After mapping the generated text into model elements, the elements that Fig. \ref{fig:AD} shows in white represent the completion elements that we obtain. The resulting completions are considered good from two different points of view. From a conceptual point of view, it fits perfectly the domain being modeled; and from a syntactic point of view, the completion suggested comply with the activity diagram syntax.

%What is represented in green correspond to elements that actually do exist in the ground truth model unlike what is represented in red which correspond to elements suggested by the engine but do not figure in the target model.
%We notice that the engine was able not only to  suggest both missing actions but also it included a decision node and provided  actions  that we can easily judge that they are relevant to the context and could be added by the user.

%\begin{figure}[h]
%\begin{center}
%\includegraphics[width=0.4\textwidth]{figures/AD-results.png}
%\caption{Example of activity diagram: Generated model}
%\label{fig:resultsAD}
%\end{center}
%\end{figure}

%Existing work focuses only on static model completion and requires a large number of training examples to be effective. With this novel idea, we show that we can improve diagram completion without the need of training or fine-tuning LMs with large sets of examples. We additionally show that few-shot prompt learning can also be applied to the completion of other domain modeling diagrams such as behavioural ones that are more constrained semantically.
\section{Preliminary evaluation and discussion}
The work presented in this NIER paper is an initial attempt to improve the completion of software models. At the current stage of our research, we have evaluated our idea on domain models represented as class diagrams from a public repository. %This is the initial step of our evaluation agenda.
\subsection{Setup}

We consider 30 domain models, selected manually from the dataset ModelSet~\cite{HernandezLopez22}.
%We avoided selecting diagrams with noisy data such as source code related details.
%The format of the  selected class diagrams is .ecore, thus we  apply model transformation techniques to extract relevant details and search for valuable information. % The Model Driven Engineering (MDE) paradigm goes one step further in the lane of this step and advocates the use of metamodels and model transformations to speed up and simplify the task.
%we study in this work the completion of domain models with related concepts or class names, with missing attributes and finally with association names.
%We follow different strategies to simulate these scenarios and we compare the obtained results using metrics among the following ones:
%To study the performance of our approach, we use the following metrics:
\begin{comment}
\begin{equation}
\footnotesize
\text{Accuracy}= \frac{\text{Correct named samples}} {\text{Total of all  samples}}
\label{accuracy}
\end{equation}
\begin{equation}
\footnotesize
\text{Recall}= \frac{\text{Number of matched elements}} {\text{Total of all relevant elements}}
\label{recall}
\end{equation}
\begin{equation}
\footnotesize
\text{Precision}= \frac{\text{Number of matched elements}} {\text{Total of all recommended elements}}
\label{precision}
\end{equation}
\end{comment}
In our experiments, we have used the GPT-3 model \emph{text-davinci-002}. Regarding the hyper-parameters, we have set the \emph{temperature} to a value between 0.70–0.90, and the \emph{maximum number of tokens} to 20 when generating suggestions for new classes (and their associations) and new attributes, and to 1 when suggesting association names.
%Although we use in our approach  a tuning free prompting, which means the engine directly generates the answers without changing the parameters of the pre-trained language model, taking into consideration  the introduced few shots only, we still  have to set other parameters. The first parameter we have to think about is the engine, this parameter  specifies the AI model employed to generate predictions. For  all the investigated tasks  in our study, we use "text-davinci-002", since  we find that it generally produces competitive or better results than the rest of engines.
%The temperature is also one of the most important settings, and in general, GPT-3 tends to select words with a larger likelihood of occurring when the temperature is lower. In our case, we avoid lower temperature because the search space should not be limited, we set it  between 0.70–0.90 as it is  most common for creative tasks.
%Another important parameter is  the maximum number of tokens that can be generated by the model. In our approach we set  it differently depending  on the task. In fact, for class names and attributes  prediction we set it to 20 because in both tasks, the aim is to generate several words and complete the missing concepts. However, for association name  prediction we set it to 1 because the aim is to precise one word.
%Regarding the different features studied in our work,
Furthermore, we have automated the completion process by automatically simulating the behaviour of a designer using our proof of concept tool.
A replication package with the code of our tool and experiments can be found on our Github repository~\cite{GithubRepo}.

\subsubsection{Class names suggestions}
To evaluate whether our approach leads to an effective suggestion of new classes, from each model \texttt{M$_i$}, we take 20\% of its elements as the already defined partial diagram \texttt{M$_i'$} % Then, we run our approach to generate completion elements and compare them with the ground-truth (i.e., \texttt{M$_i$}) to classify them as relevant or irrelevant suggestions.
% We  have first selected randomly from every full model   a number of concepts corresponding to 20$\%$ of the total concepts.
and we simulated an incremental design process starting from \texttt{M$_i'$}. First, we performed a first round (R1) of completion suggestions, then  we validated the suggested results  manually adding to the model under-construction \texttt{M$_i$'} those which were semantically equivalent to those elements in \texttt{M$_i$}. The accepted elements were included in the partial model \texttt{M$_i'$}, resulting in the partial model \texttt{M$_i''$}. Then, we performed a similar second round (R2) starting from the partial model \texttt{M$_i''$}.

\subsubsection{Attributes suggestion}
To assess the effectiveness of our approach when suggesting new attributes within a class, we have selected randomly 212 classes from our dataset, have removed 75\% of their attributes and have generated attribute completions for them. Note that this means that for classes with three or less attributes, we removed all of them, which was the case for most of the classes.
%To do so, we only perform one step because we notice that the number of attributes in  most of the classes in our validation dataset is not high (average of 2).
Once we obtained the completion suggestions from our engine, we manually approved those which are either exact matches or semantically equivalent elements to those in the ground-truth model.

%\lola{probably we should move this to Section \ref{sec:staticDiagCompletion}, because this explanation is part of our approach.} \lola{I don't remember exactly how we did it in the end and, with this info, it is not completely clear to me. When recommending attributes for a class, we used: the class name, the existing attributes in the class, and the package? or did we add more info (i.e., other classes' names)? It would be great to be more concrete}

\subsubsection{Association names  suggestion}
We finally evaluate whether our approach is able to suggest meaningful association names.
We extract from our dataset 40 pairs of concepts, where each pair contains the names of two associated classes.
Then, we query 3 times our engine to suggest a name for each association, each attempt with  the same prompt but a different temperature. % and we use a frequency based ranking function to suggest the final association name.
We validated manually the output of our engine  and approved those which are either exact matches or semantically equivalent to those in the ground-truth model.
\vspace{-1mm}

\subsection{Results}

%\lola{Same as before, it should be moved to Section~\ref{sec:staticDiagCompletion} and I am missing more details to fully understand how we have done it... Meriem, if you add these details I can go through the text tomorrow.}

%\lola{Since we are moving the text above to the previous section, the content of this section should be how we have evaluated our approach (as explained for concept suggestion). This is, how many class diagrams we used for attributes and association suggestions respectively, hyperparameters, etc. }

%\lola{Somewhere we should also write the formulas we have used to calculate the precision and recall}
\vspace{-1mm}

\subsubsection{Class names suggestion}

As explained previously, we collect results for two successive steps. Table \ref{tab:tableConceptSuggestions} summarizes the precision and recall metrics for both steps. We observed that the \textit{Recall} improved from R1 to R2 while the \textit{Precision} decreased slightly.
This is due to the fact that the number of correctly suggested elements  increases from one round to the next, while the number of incorrect elements increases too.
\vspace{-4mm}
\begin{center}
\scriptsize
\begin{table}[H]
\centering
\begin{tabular}{ || c| c | c |  c | c|| }
 \hline
 & Precision R1 & Precision R2 & Recall R1 & Recall R2 \\
  \hline
 avg & 0.57 & 0.56 & 0.29 & 0.45 \\
  \hline
 std & 0.26	& 0.24 & 0.18 &	0.25 \\
\hline
\end{tabular}
\label{tab:tableConceptSuggestions}
\caption{Results evaluating class names prediction}
\vspace{-10mm}
\end{table}
\end{center}
We also observed that domain models that resulted in the best results (recall 0.8 to 1) were addressing very common domain/topics used by humans in natural language such as \texttt{Bank, University} and \texttt{library} system. Yet, domains with noisy information and non understandable data, such as a model whose package name was \texttt{AUni}, resulted into poor results (0 to 0.1 recall).

\subsubsection{Attributes suggestion}
%The recall metric, defined as the ratio of the ground-truth attributes being found in the Top-N recommended items,  was used to study the performance of this strategy.

For attribute suggestions, we only evaluate the recall because most of the classes in these selected domain models contain a very few number of attributes, thus our interest is to check whether we are able  to obtain these missing attributes.
\begin{comment}
\begin{center}
\scriptsize
\begin{tabular}{ || c| c || }
 \hline
  & Recall \\
  \hline
  avg & 0.7 \\
    \hline
  svd & 0.4 \\

 \hline
\end{tabular}
\end{center}
\end{comment}
The average recall is \texttt{0.7} with a standard deviation is \texttt{0.4}, which can be considered a promising result. %However the std shows a high variability in the set of our data.

\subsubsection{Association names suggestion}
An interesting metric to evaluate these suggestions is the accuracy, defined as the ratio of correctly predicted association names w.r.t. the total suggestions. This is, unlike before, we are no longer interested in recognizing the relevant elements, but checking how many times the engine was correct. We have obtained an accuracy of \texttt{0.64}, which also seems promising.

\begin{comment}
\scriptsize
\begin{tabular}{ || c|| }
 \hline
Accuracy  \\

 \hline
 0.64 \\

 \hline
\end{tabular}
\end{comment}

%\lola{No avg and std for the association names predictions?}

%\lola{I think we should explain why we're only showing the recall for attributes and association names suggestions and not the precision (I don't remember why, but if we don't explain it, it looks strange.}

\section{Conclusion and Discussion}
We proposed a novel approach based on few-shot prompt learning to enable large language models to solve completion tasks in modeling activities. We reformulate model completion as a semantic mapping problem that consists, firstly, in transforming modeling formalism elements into meaningful patterns of sequences of tokens to create prompts with learning shots. Then, we exploit the ability of LLMs to complete partial sequences following the specified patterns to recover elements that can be used for the completion. Those elements are transformed into constructs conforming to the modeling language syntax and suggested to the designers.  
Although many research contributions were proposed to solve model completion problems, we do believe that none of them can be effectively used in a real setting, because of the resources needed, i.e., large training datasets, and the limited performance they offer. We do believe, however, that our approach can be effective when modeling both static and dynamic diagrams for two main reasons. Firstly, it does not require to pre-train or fine tune language models on specific tasks or domain. Secondly, the used LLMs are trained on a huge volume of data, which makes it generalizable to many domains and different concept natures and relationships. 

Although our approach shows promising results, it is still a first attempt and there is room for improvement. Indeed, when defining a prompt, the elements of the already-defined partial diagrams have a great influence on the accuracy of the suggested token. A calibration study is still necessary to determine the boundaries of the provided existing context to have the best suggestions. For example, when we added systematically the package name in the pattern, the results improved considerably, but we cannot determine whether this observation is valid only on the used benchmark.
Another consideration that has to be studied is the use of non-natural language elements such as symbols and digits. In our experiments, their existance generated poor results as these elements are rarely present in the data used for training LLM. We believe that a more sophisticated mapping of those elements would considerably improve the results.
%Another aspect worth investigating is the number of shots necessary to generate optimal results.
Using LLMs proves to be efficient in modeling formalism that rely heavily on natural language identifiers. However, other modeling languages such as Petri nets are definitely difficult to handle as they involve modeling elements that cannot be captured by LLMs. 

In summary, the proposed approach, although simple to implement, is powerful and showed promising results. Those results are possibly even better if we consider the fact that the boundaries of a domain are broader than the diagrams we considered as ground truth in our experiments. In fact, some of the suggested elements, although absent in the considered diagrams may be relevant for the considered domain. To asses such a claim, we plan to conduct a user study to better assess the correctness of the suggestions, but also the usefulness of the completion for the domain modelers.

\bibliographystyle{IEEEtran}
\bibliography{references}

\end{document}